\documentclass[twoside,fleqn]{article}
\usepackage{latexsym}
\usepackage{espcrc2}
\usepackage{graphicx}
\usepackage{epsf}

\newcommand{\AmS}{{\protect\the\textfont2
  A\kern-.1667em\lower.5ex\hbox{M}\kern-.125emS}}

\newlength{\numlen}

\newcommand{\nr}[1]{(\ref{#1})}
\newcommand{\fr}[2]{{\frac{#1}{#2}}}

\newcommand{\be}{\begin{equation}}
\newcommand{\ee}{\end{equation}}
\newcommand{\ba}{\begin{eqnarray}}
\newcommand{\ea}{\end{eqnarray}}
\newcommand{\bi}{\begin{itemize}}
\newcommand{\ei}{\end{itemize}}

\def\lsi{\raise0.3ex
\hbox{$<$\kern-0.75em\raise-1.1ex\hbox{$\sim$}}}
\def\gsi{\raise0.3ex
\hbox{$>$\kern-0.75em\raise-1.1ex\hbox{$\sim$}}}
\newcommand{\lsim}{\mathop{\lsi}}

\newcommand{\fig}{Fig.~}

\title{%
Critical behaviour of the Ginzburg-Landau model in the type II region
}

\author{%
K. Kajantie\address{Department of Physics, P.O.Box 64, 
FIN-00014 University of Helsinki, Finland},
M. Laine\address{Theory Division, CERN, CH-1211 Geneva 23, Switzerland},
T. Neuhaus\address{Finkenweg 15, D-33824 Werther, Germany},
A. Rajantie\address{DAMTP, University of Cambridge, 
Cambridge CB3 0WA, United Kingdom}
and
K. Rummukainen\address{NORDITA, Blegdamsvej 17, 
DK-2100 Copenhagen \O, Denmark}%
\hfill\raisebox{20mm}[0mm][0mm]{\makebox[0mm][r]{NORDITA-2001/37 HE, October 2001}}%
}
\begin{document}

\begin{abstract}
We study the critical behaviour of the three-dimensional U(1) gauge+Higgs
theory (Ginzburg-Landau model) at large scalar self-coupling $\lambda$
(``type II region'') by measuring various correlation lengths as well as 
the Abrikosov-Nielsen-Olesen vortex tension.  We identify different scaling
regions as the transition is approached from below, and carry out detailed
comparisons with the criticality of the 3d O(2) symmetric scalar theory.
\end{abstract}

\maketitle


The three-dimensional (3d) 
U(1) gauge+Higgs theory (Ginzburg-Landau (GL) model, scalar
electrodynamics) is an effective theory for phase transitions in
superconductors, liquid crystals, and possibly cosmology.  The Lagrangian is
\be
  {\cal L}_{GL} =  \fr14 F_{ij}^2 + |D_i\phi|^2 
	+ m_3^2 |\phi|^2 + \lambda_3 |\phi|^4 \,.
\label{lagrangian}
\ee
Here $\phi$ is a complex scalar field, $F_{ij} = \partial_i A_j - \partial_j
A_i$, and $D_i = \partial_i + ie_3 A_i$.  The parameters
$m_3$, $e_3^2$, $\lambda_3$ have the dimension GeV.
Physics then depends on the dimensionless ratios
\[
  x \equiv \lambda_3/e_3^2, ~~~~
  y \equiv m_3^2/e_3^4.
\]

\begin{figure}[bt]
\centerline{\epsfxsize=7cm\epsfbox{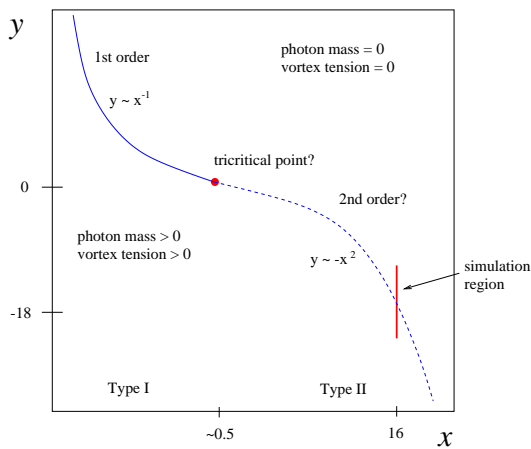}}
\vspace{-1cm}
\caption[a]{The schematic phase diagram of the GL model.}
\label{fig:schematic}
\vspace*{-5mm}
\end{figure}

The GL model has no local order 
parameters.  However, there are non-local quantities, like the
photon mass $m_\gamma$, and the vortex tension $T$ \cite{vortextension},  
which vanish in the symmetric and are non-zero in the broken phase.
The schematic phase diagram is shown in \fig\ref{fig:schematic}.
At small $x$ (type I region) the transition is of the first order, 
while at large $x$ (type II region) the transition is 
believed to be of the second order, with
a (presumably) tricritical point in between, at $x \approx 0.3$ \cite{hove}.

In this work we study the phase transition deep in the type
II region, at $x = \lambda_3/e_3^2 = 16$, shown as a thick vertical
line in \fig\ref{fig:schematic}, and compare the scaling behaviour with
that of
the 3d XY model. 
The universality
class of the transition in the type II domain has not been previously
resolved unambiguously with lattice simulations.  

The scaling behaviour of the
system is determined by the longest correlation lengths, 
or the lightest ``masses''. We shall
consider the photon and scalar masses $m_\gamma$ and $m_s$ as functions of
$\tau \equiv (y-y_c)$.  The relation between masses varies strongly as
$\tau$ is varied; consequently, 
in the broken phase, one can argue that there are three
distinct possibilities:

\noindent
{\em 1. Mean field:}~ The textbook case,
valid if $\tau \ll 0$, leading to
$m_\gamma \sim (|\tau|/x)^{1/2} \ll  m_s \sim (2|\tau|)^{1/2}$.

\noindent
{\em 2. XY scaling:}~ Here
scaling is dominated by the dynamics of the scalar field; indeed, since 
$\lambda_3 \gg e_3^2$, it is not unreasonable for this to happen.
Such scaling is supported by RG arguments \cite{rg},
and has been observed experimentally \cite{Kamal}.
In this domain 
$
m_s \sim |\tau|^{\nu_{\rm xy}}, ~ m_\gamma \sim |\tau|^{\nu_{\rm xy}/2},
$
with $\nu_{\rm xy}\approx 0.67$,
and we expect $m_s < 2 m_\gamma$, 
otherwise the scalar would decay into two photons.  

However, we do not expect this scaling to remain
valid very close to the critical point:
the XY model has a massless Goldstone at $\tau \le 0$,
whereas the GL model has a massless photon only when $\tau \ge 0$.
Moreover, due to logarithmic
confinement in 3d U(1) gauge theory, one can expect the scalar 
to remain massive at the transition point.

\noindent
{\em 3. Inverted XY scaling:}~ Here the GL
``temperature'' parameter $\tau$ is supposed to map to the {\em inverted}
temperature ($-\tau$) of the 3d XY model, and the broken/symmetric
phases of the GL model are supposed to 
correspond to the symmetric/broken phases of the
XY model.  This scenario is supported by duality arguments \cite{duality}.
When $\tau < 0$ the GL model in (2+1)d has two massive photon polarisation states, which
become critical at $\tau =0$. At $\tau > 0$, these d.o.f's become 
the massless photon, and a massive vector ``resonance''.
These counts match exactly 
the two massive scalars in the symmetric phase 
of the XY model, which, in the broken phase, become the massless Goldstone
and the massive radial mode.

Identification of the photon with the critical degree of freedom of the
XY model suggests
$m_\gamma \sim |\tau|^{\nu_{\rm xy}}$;
however, other estimates in the literature suggest
$m_\gamma \sim |\tau|^{0.5 \ldots 0.67}$ \cite{duality,Herbut}.
The scalar remains massive, since all diverging 
d.o.f's have been accounted for. (Alternatively, the scalar
mass could vanish, but with an exponent smaller than the photon one). 

Duality arguments also relate the Abrikosov-Nielsen-Olesen vortex
tension $T$ to the scalar mass in the dual theory,
giving $T \sim |\tau|^{\nu_{\rm xy}}.$

We use the standard lattice discretization of Eq.~\nr{lagrangian},
with a non-compact U(1) gauge action.  For details, see
\cite{vortextension}.  We choose a fixed lattice spacing ($e_3^2 a = 1$);
observing the universal behaviour does not require taking 
the continuum limit.  

Unfortunately, in the GL model 
there is no gauge invariant local observable which
would correspond to magnetization.  Vortex
percolation has often been used to probe the critical 
behaviour~\cite{nguyen}; however, on the lattice this is not 
rigorous~\cite{percolation}.  We shall here only measure
quantities such as the specific heat, scalar and photon
correlation lengths, and the vortex tension. 

In this paper we report on correlation lengths.
In the GL model, these are measured using the following operators
(for concreteness, measurements taken along the $x_3$-direction):

\begin{tabular}{ll}
{\em Scalar }              & $ S = \phi^*\phi$, \\
{\em Photon (plaquette) }  & $ B_3 = \epsilon_{123} F_{12} = \Box_{12}$, \\
{\em Vector }              & $ V_i = \mbox{Im\,}\phi^* D_i \phi $\,.
\end{tabular}
The vector and photon operators are measured at finite ``transverse
momentum'' to, say, the $x_1$-direction.  Additionally, each channel
contains states with multiple photons, complicating the
scalar channel measurement in particular. 
In the symmetric phase the scalar is
always a resonance, decaying into two or more photons.  
For each operator we use several levels of {\em blocking + smearing}
to reduce noise, and we diagonalize the full cross-correlation matrix
between the blocked operators, as described, e.g.,~in~\cite{op}. 

\begin{figure}[bt]
\centerline{\epsfxsize=6cm\epsfbox{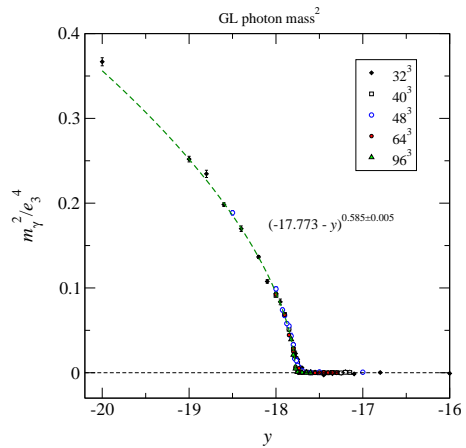}}
\vspace{-1cm}
\caption[a]{The photon (mass)$^2$ in the GL model.}
\label{fig:photonmass}
\vspace*{-5mm}
\end{figure}

In \fig\ref{fig:photonmass} the photon mass squared is shown.
The critical exponent is $2\nu
\approx 0.59$. Inverted XY scaling is
usually thought to imply a larger exponent,
$1.0\lsim 2\nu \lsim 2\nu_{\rm xy} \approx 1.34$ (see above). 
Thus, we conclude that inverted XY scaling, 
at least in this form,
is not valid for these $\tau=y-y_c$.

\begin{figure}[bt]
\centerline{\epsfxsize=6cm\epsfbox{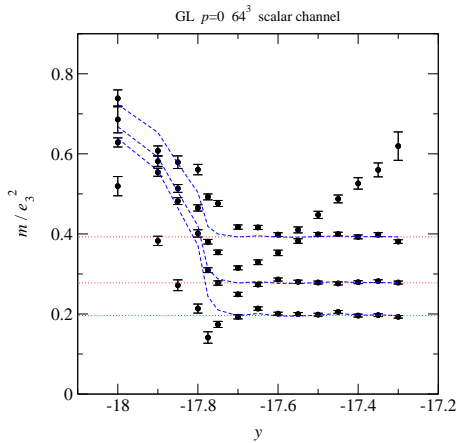}}
\vspace{-1cm}
\caption[a]{The four lowest mass states in the scalar channel.  
Horizontal dotted lines: masses of the three lowest two-photon states,
with (single photon) $m_\gamma=0$.  Dashed lines: masses of two-photon
states, using $m_\gamma$ from \fig\ref{fig:photonmass} as input.}
\label{fig:resonance}
\vspace*{-5mm}
\end{figure}

\begin{figure}[bt]
\centerline{\epsfxsize=6cm\epsfbox{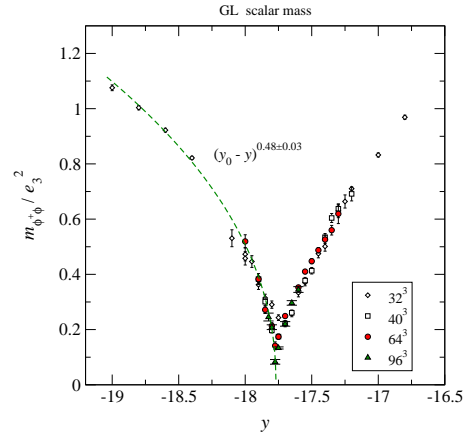}}
\vspace{-1cm}
\caption[a]{The scalar resonance mass.}
\label{fig:scalarmass}
\vspace*{-5mm}
\end{figure}

In the scalar channel we include up to four lowest two-photon states
at zero total transverse momentum in the cross-correlation analysis,
requiring diagonalization of matrices up to $14^2$, 
and improve thus on earlier work at $x=2$~\cite{old}.  
In \fig\ref{fig:resonance} we show measurements
on $64^3$ lattices.  Three of the states clearly
correspond to two photons, 
and the scalar ($\phi^*\phi$) is a resonance in the 
symmetric phase.

The scalar (resonance) mass is shown
separately in \fig\ref{fig:scalarmass}.  
Approaching $y_c$ from the broken phase, the mass appears to
vanish with a critical exponent $m_s \propto
|\tau|^{0.48}$.  
The value does not agree well with ``standard'' XY scaling, 
$\nu_{\rm xy}\approx 0.67$, but
it is marginally compatible with the
inverted XY scaling scenario, where the scalar mass
should either remain finite at $y_c$, or at least have a 
critical exponent smaller
than $\nu_{\rm xy}$.
%
%

In conclusion, the 
photon correlation length does
not display (at least with the signatures commonly suggested)
inverted XY scaling, as
predicted by duality arguments. 
In addition, the critical
exponent for the vortex tension $T$ is several standard deviations
off from the inverted XY
scaling prediction (to be reported elsewhere).

Of course, one possible reason
for the discrepancy is that at $x=16$ the scaling window may 
be an extremely
narrow band around $y_c$, rendering it almost impossible to observe
the true scaling in Monte Carlo simulations.

\end{document}